\documentstyle[preprint,aps]{revtex}
\begin{document}
\preprint{UM-P-93/52, OZ-93/14}
\draft
\title{Constraints on the anomalous $WW\gamma$ couplings from
$b \rightarrow s \gamma$ }
\author{Xiao-Gang He and Bruce McKellar}
\address{Research Center for High Energy Physics\\
School of Physics\\
University of Melbourne \\
Parkville, Vic. 3052 Australia}
\date{May, 1993}
\maketitle
\begin{abstract}
We study contributions to $b \rightarrow s \gamma$ from anomalous $WW\gamma$
interactions. Although these anomalous interactions are not renormalizable,
the contributions are cut-off independent.
Using recent results from the CLEO collaboration on inclusive
radiative B decays, we obtain bounds for the anomalous CP
conserving and CP violating couplings.
The constraints on the CP conserving couplings
are comparable with or better than constraints from other experiments.
\end{abstract}
\newpage
The Minimal Standard Model (MSM) of electroweak interactions is in very good
agreement with present experimental data. However, there are still many
questions unanswered. The structure of the MSM has to be tested in fine
detail. One particular interesting point is to find out the structure of
the self-interaction of the electroweak bosons. Study of this will help us to
establish whether the weak bosons are gauge particles with interactions
predicted by the MSM, or gauge particles of
some extensions of the MSM which predict different
interactions at loop levels, or even non-gauge particles whose
self-interactions at low energies are described by effective interactions.
In general there will be more self-interaction terms than the tree level
MSM terms - these additional terms are refered to as the anomalous couplings.
It is important to find out experimentally what
are the allowed regions for these anomalous couplings.
In this paper we will study
constraints on the anomalous $WW\gamma$ couplings
using experimental data from $ b \rightarrow s \gamma$.

The CLEO
collabortion has recently observed\cite{cleo}
 the exclusive decay $B \rightarrow
K^* \gamma$ with a branching ratio $ (4.5 \pm 1.5 \pm 0.9) \times 10^{-5}$
and placed an upper limit on the inclusive quark-level process $b
\rightarrow s \gamma$ of $B(b \rightarrow s \gamma) < 5.4\times 10^{-4}$
at 95\% CL. In the MSM, $b \rightarrow s \gamma$  is induced at the one
loop level \cite{inc,excl}.
The inclusive branching ratio is typically about
$2\times 10^{-4}$, which is consistent
with the experiment. The theoretical prediction for the exclusive decay
$B \rightarrow K^* \gamma$ depends on hadronic form factors which are not
well determined at present. The ratio $Br(B\rightarrow K^*\gamma)
/Br(b\rightarrow s \gamma)$ is estimated to be between 0.4 to 0.04\cite{excl}.
The MSM is not in conflict witht the experimental data. Due to the large
uncertainty associated with the hadronic form factors for $B\rightarrow
K^* \gamma$, in our study we will only
consider constraints on the anomalous $WW\gamma$ couplings from the upper
bound on $Br(b\rightarrow s \gamma)$.

The most general form, invariant under $U(1)_{em}$, for the anomalous
$WW\gamma$
interactions can be parametrized as
\begin{eqnarray}
L &=& i\kappa W^+_\mu W^-_\nu F^{\mu\nu} + i {\lambda\over M_W^2}
W^+_{\sigma\rho}W^{-\rho\delta} F^\sigma_\delta\nonumber\\
&+&  i\tilde \kappa W^+_\mu W^-_\nu \tilde F_{\mu\nu}
+ i{\tilde\lambda \over M_W^2}W^+_{\sigma\rho}W^{-\rho\delta}\tilde
F^\sigma_\delta\;,
\end{eqnarray}
where $W^\pm_\mu$ are the W-boson fields, $W_{\mu\nu}$ and $F_{\mu\nu}$ are
the W-boson and photon field strengths respectively, and $\tilde F_{\mu\nu}
={1\over 2}\epsilon_{\mu\nu\alpha\beta}F^{\alpha\beta}$.
The terms proportional to $\kappa$ and $\lambda$ are CP conserving and $\tilde
\kappa$ and $\tilde\lambda$ are CP violating. The contributions to $b
\rightarrow s\gamma$ from the $\kappa$ and $\lambda$ terms have been studied
by Chia\cite{chia} and Peterson\cite{pet}.
Here we will consider all the contributions and give
up-dated constraints on $\kappa$ and $\lambda$, and new constraints on
$\tilde \kappa$ and $\tilde \lambda$.

The process $b \rightarrow s \gamma$ is induced at the one loop level. The
effective Hamiltonian is given by
\begin{eqnarray}
H_{eff} &=& i{g^2\over 2}\epsilon^\mu V_{tb}V_{ts}^* \bar s
\gamma_\alpha \gamma_\nu \gamma_\beta {1-\gamma_5 \over 2} b\nonumber\\
&\times&\int {dk^4\over (2\pi)^4} {k^\nu (g^{\alpha\alpha'} -
{k^{+\alpha} k^{+\alpha'}\over M_W^2})
(g^{\beta\beta'}-{k^{-\beta}k^{-\beta'}\over M_W^2})\Gamma_
{\mu\alpha'\beta'}(q, k^+, k^-)
\over (k^2-m_t^2)((p-k)^2-M_W^2)
((p'-k)^2-M_W^2)}\;,
\end{eqnarray}
where
\begin{eqnarray}
\Gamma_{\mu\alpha\beta}(q, k^+, k^-)&=&
-\kappa (g_{\alpha\mu}q_{\beta}-g_{\beta\mu}q_{\alpha})
-\tilde \kappa \epsilon_{\mu\alpha\beta\rho}q^\rho \nonumber\\
&+& {\lambda\over M_W^2}(g_\alpha^\rho k^{+\delta} - g_\alpha^\delta
k^{+\rho}) (g_\beta^\delta k^{-\sigma} - g_\beta^\sigma k^{-\delta})
(g_\rho^\mu q^\sigma - g^\sigma_\mu q^\rho)\nonumber\\
&+&{\tilde \lambda\over M_W^2}(g_\alpha^\rho k^{+\delta} - g_\alpha^\delta
k^{+\rho})(g_\beta^\delta k^{-\sigma} -g_\beta^\sigma k^{-\delta})
\epsilon_{\rho\sigma\mu\tau}q^\tau\;,\nonumber
\end{eqnarray}
where $k$, $p$, and $p'$ are the internal,
b-quark and s-quark momentum respectively,
$q = p' - p$, $k^+ = p - k$ and $k^- = k -p'$, and $\epsilon^\mu$ is the photon
polarization. Performing the standard Feynman parametrization, we have
\begin{eqnarray}
H_{eff} &=& ig^2\epsilon^\mu V_{tb}V_{ts}^*\bar s
\gamma_\alpha\gamma_\nu\gamma_\beta {1-\gamma_5\over 2}b\nonumber\\
&\times& \int^1_0dx \int^{1-x}_0dy \int {d^4k\over (2\pi)^4}
{k^\nu(g^{\alpha\alpha'} - {k^{+\alpha}k^{+\alpha'}\over M_W^2})
(g^{\beta\beta'} - {k^{-\beta}k^{-\beta'}\over M_W^2})\Gamma_{\mu\alpha'\beta'}
(q,k^+,k^-)\over
(k^2 - 2k\cdot(xp+yp')-(m_t^2 +(M_W^2 -m_t^2)(x+y))^3}\;.\nonumber\\
\end{eqnarray}
Substituting $k' = k-(xp+yp')$ into eq.(3), we find only terms quadratic in
$k'$ contribute to $b \rightarrow s \gamma$. Higher order terms vanish
and lower order terms are supressed by factors like ${m_b^2\over M_W^2}$
or ${m_s^2\over M_W^2}$, and can be safely neglected.
We find the effective Hamiltonian $H^\gamma_{eff}$
responsible for $b\rightarrow s \gamma$ to be
\begin{eqnarray}
H^\gamma_{eff} &=& {g^2\over 2M_W^2}V_{tb}V_{ts}^*m_b\bar s
\sigma_{\mu\nu} \epsilon^\mu q^\nu {1+\gamma_5\over 2} b\nonumber\\
&\times&\int^1_0dx\int^{1-x}_0 dy \int {dk^4\over (2\pi)^4}
{(\kappa-i\tilde \kappa) (2(x+y)-k'^2) +(\lambda -i \tilde \lambda)
k'^2(2-3(x+y)) \over \beta_t + (1-\beta_t)(x+y)}\;,
\end{eqnarray}
where $\beta_t = {m_t^2\over M_W^2}$.
In the above we have neglected terms proportional to $m_s$ and have used the
equation of motion and the
identities: $\bar s \sigma_{\mu\nu}\gamma_5 b = -{\epsilon_{\mu\nu\alpha\beta}
\over 2}\bar s \gamma^\alpha\gamma^\beta b$, and $\bar s \sigma_{\mu\nu} b =
-{\epsilon_{\mu\nu\alpha\beta}\over 2}\bar s \gamma^\alpha\gamma^\beta
\gamma_5b$. After integrating over $k'$, we obtain
\begin{eqnarray}
H^\gamma_{eff} &=&i{g^2\over 32\pi^2M_W^2}
V_{tb}V_{ts}^*m_b\bar s \sigma_{\mu\nu} \epsilon^\mu q^\nu
{1+\gamma_5\over 2}b\nonumber\\
&\times&[(\kappa-i\tilde\kappa)(I_1(\beta_t) -I_1(0))
+(\lambda -i\tilde\lambda)(I_2(\beta_t)-I_2(0))]\;,
\end{eqnarray}
with
\begin{eqnarray}
I_1(x)&=&\int^1_0dx\int^{1-x}_0dy [{x+y\over \beta_t
+(1-\beta_t)(x+y)} - \ln (\beta_t+(1-\beta_t)(x+y))]\nonumber\\
I_2(x)&=&\int^1_0dx\int^{1-x}_0dy(2-3(x+y))\ln (\beta_t+
(1-\beta_t)(x+y))\;.
\end{eqnarray}
Here we have used the GIM mechanism to cancel out terms which do not depend
on internal quark masses.
The final expression for $H^\gamma_{eff}$ is given by
\begin{eqnarray}
H^\gamma_{eff} = i G_2 m_b \bar s
\sigma_{\mu\nu}\epsilon^\mu q^\nu {1+\gamma_5 \over 2} b\;,
\end{eqnarray}
where
\begin{eqnarray}
G_2 = {G_F\over \sqrt{2}} {e\over 4 \pi^2} V_{tb}V_{ts}^*
G_A({m_t^2\over M_W^2})\;,
\end{eqnarray}
with $G_A(x)$ given by
\begin{eqnarray}
G_A(x) &=& ({\kappa\over e} - i{\tilde\kappa\over e})({x\over (1-x)^2}
+ {x^2(3-x)lnx\over 2(1-x)^3})\nonumber\\
&&- ({\lambda\over e} - i{\tilde\lambda \over e})
({x(1+x)\over 2(1-x)^2} + {x^2lnx\over (1-x)^3})\;.
\end{eqnarray}
We emphasise that the result is finite and that we do not need the use of
cut-offs to analyze our results.

Combining the contribution from the MSM and using the leading QCD correction
obtained by Grintein, Springer and Wise in Ref.\cite{inc},
we obtain the total contribution to $b \rightarrow s \gamma$ of the form of
 eq.(7), but with $G_2$
replaced by $G^t_2$. Here $G^t_2$ is given by
\begin{eqnarray}
G^t_2 &=& i{G_F\over \sqrt{2}}{e\over 4\pi^2} V_{tb}V_{ts}^*
C_7({m_t^2\over M_W^2}) \;,\nonumber\\
\nonumber\\
C_7(x) &=& \left( {\alpha_s(M_W)\over \alpha_s(m_b)}\right)^{16/23}
\Big( C'_7(x)+G_A(x) - {8\over 3}C'_8(x)\left[ 1-
\left( {\alpha_s(m_b)\over \alpha_s(M_W)}
\right) ^{2/23}\right] \nonumber\\
&&+ {464\over 513}\left[ 1-\left({\alpha_s(m_b)\over \alpha_s(M_W)}\right)
^{19/23}\right] \Big) \;,
\end{eqnarray}
and
\begin{eqnarray}
C'_7 (x) &=& {x\over (1-x)^3}({2\over 3}x^2+{5\over12}x - {7\over 12}
+{({3\over 2}x^2 - x)lnx\over  1-x})\nonumber\\
C'_8(x) &=& {x\over 2(1-x)^3}({1\over 2} x^2 - {5\over 2}x - 1 -
{3xlnx\over 1-x})\;.
\end{eqnarray}
The decay width for $b \rightarrow s \gamma$ is given by
\begin{equation}
\Gamma(b\rightarrow s \gamma) = {|G_2|^2m_b^5\over 16\pi}\;.
\end{equation}

To obtain $Br(b \rightarrow s \gamma)$,
we use the latest data on the semileptonic branching ratio\cite{dre}
 $Br(b \rightarrow
X_c e\nu) = 0.108$ to scale the inclusive $b \rightarrow s \gamma$ rate. This
proceedue removes the uncertainties associated with $(m_b)^5$ and KM factors.
We have
\begin{equation}
Br(b\rightarrow s \gamma) = {|V_{tb}V_{ts}^*|^2\over |V_{cb}|^2}
Br(b\rightarrow X_c e\nu){3\alpha_{em}\over
2\pi \rho \eta}|C_7({m_t^2\over M_W^2})|^2\;,
\end{equation}
where the phase space
factor $\rho = 1 - 8 r^2 +8r^6-r^8-24r^4\ln r$ with $r=m_c/m_b$
and the QCD correction factor $\eta = 1- 2f(r,0,0)\alpha_s(m_b)/3\pi$ with
$f(r,0,0) = 2.41$\cite{cab}. In our numerical analysis,
we will use\cite{part}: $\alpha_s(M_W)=
0.105$, $m_b = 5 GeV$,
$m_c = 1.5 GeV$, $|V_{tb}V_{ts}^*|/|V_{cb}| \approx 1$ and let the top quark
mass $m_t$ and the anomalous couplings vary.

The contributions from the $\kappa$ and $\lambda$ terms can either increase or
decrease the MSM prediction for $Br(b\rightarrow s \gamma)$
depending on the ranges and the
signs of the parameters. If we set the signs of $\kappa$ and $\lambda$ to be
positive, the contribution from the $\lambda$ term has the same sign as that of
the MSM, while the contribution from the $\kappa$ term
has opposite sign.
It is confusing
to do the complete analysis if we keep $\kappa$, $\tilde \kappa$,
$\lambda$, and $\tilde \lambda$ arbitrary. For simplicity
we will consider constraints on individual anomalous coupling;
that is, we let only one anomalous
coupling to be non-zero when carrying out the analysis.
The resulting bounds on the anomalous couplings at the 
are given in Table 1. The constraint
on $\kappa$ is the tightest, for example, for $m_t = 150 GeV$, $\kappa/e$ is
constrained to be between $ 2.5$ and  $-0.44$. This bound is better than that
obtained from the muon anomalous magnetic dipole moment\cite{mag}.
Our bound restrict $\kappa$ to
a range than that derived from an analysis of the experimental data at CDF,
which is  $3.7 >\kappa/e >-2.6$ at 96\%\cite{sam},
and if $\kappa$ is negative this constraint is better.
The constraint on $\lambda$ is weaker, for $m_t
= 150 GeV$, $\lambda/e$ is restricted to be between $-7.2$ and $1.3$.
This bound is comparable with
the constraint from the muon anomalous magnetic dipole moment.
All constraints become tighter when $m_t$ is increased. It is interesting to
note that if the branching ratio of $b \rightarrow s\gamma$ turns out to
be significantly smaller than $2\times 10^{-4}$, it may be an indication for
non-zero anomalous $WW\gamma$ coupling. Complete
cancellation between the MSM and the anomalous contributions can occur for
certain values of $\kappa$ and $\lambda$. In
Tables 2.and 3., we give $Br(b\rightarrow s\gamma)$ as functions of $\kappa$
and
$\lambda$ for $m_t = 150 GeV$. From Tables 2. and 3., we see that if a
lower bound for $Br(b\rightarrow s\gamma)$ will be well established, one can
exclude some regions in the allowed parameter space given in Table 1.

Due to phase differences
in the amplitudes, the contributions from the $\tilde \kappa$ and $\tilde
\lambda$ terms do not interfere
with the MSM contribution
and thus appear only quadratically in the modified expression for
$Br(b\rightarrow s\gamma)$. Therefore non-zero $\tilde \kappa$ and $\tilde
\lambda$ can only increase the MSM prediction for $Br(b\rightarrow s\gamma)$.
However cancellation can occur between the
contributions from the $\tilde \kappa$ and $\tilde \lambda$
terms. Again we will analyze the constraints on the couplings by  keeping only
one non-zero. We find that the typical
constraints on $\tilde \kappa/e$ and $\tilde \lambda/e$ are of order one.
The results
are shown in Table 1. These bounds are weaker than the ones obtained
from the experimental upper bound on the neutron and electron electric dipole
moment\cite{ele}; however these bounds are cut-off dependent, and the present
bounds are not.

The above analysis has neglected a number of theoretical uncertainties.
There are uncertainties in the numerical factors for the QCD
corrections\cite{qcd}. However these corrections are small, at most a few
percent. There is also uncertainty from the KM matrix elements\cite{part},
but this uncertainty is again small. The biggest uncertainty is in
the phase space factor $\rho$. Within the allowed values
for $m_c/m_b$, $\rho$ can
vary between 0.41 to 0.65. Our choice of $m_c/m_b = 0.3$ correcponds to $\rho
=0.52$. Had we used a smaller value for $\rho$, we would have obtained stronger
constraints. The uncertainty due to possible variations in $\rho$
can have a factor of two effect on
the constraints of the anomalous couplings.

It is possible to carry out an
analysis keeping all anomalous couplings arbitrary. This exercise will obtain
a correlated allowed range in the parameter space. However we do not think
this will provide substantial new information. In fact if one chooses values
for $\kappa$ and $\lambda$ such that their contributions to $b\rightarrow
s\gamma$ cancel out, any value for $\kappa$ is allowed. Similar situation
happens to $\tilde \kappa$ and $\tilde \lambda$.

In summary, we have obtained constraints on the anomalous $WW\gamma$
couplings by using the recent results from CLEO collaboration on the
inclusive radiative B decays.  The constraints on the CP violating
couplings are weaker than that obtained from the neutron and electron
electric dipole moment. For CP conserving couplings, the constraints obtained
in this paper are comparable with or better than other constraints. Improved
experimental data on $b\rightarrow s\gamma$, especially well established
lower bound on the branching ratio can further constrain the anomalous
couplings.

\acknowledgments
This work was supported in part by the Australian Research Council.
\newpage
Note added.- After the completion of this paperr we became aware of a paper
by T. Rizzo\cite{riz} in which a similar analysis for $\kappa$ and $\lambda$
has been carried out. In addition to the analysis in our paper,
in Ref.\cite{riz} information obtained from hadron colliders are also used, and
the correlated allowed regions
for $\kappa$ and $\lambda$ are
presented.  Our results agree with that obtained by Rizzo.
The results on $\tilde \kappa$ and $\tilde \lambda$ are, however,
new. Note that there
is a sign difference in the definition of the parameter $\kappa$.

\begin{table}
\caption{The constraints for the anomalous $WW\gamma$ couplings.}
\begin{tabular}{|c|c|c|c|c|c|}
$m_t$(GeV) & 100 &125 &150 &175 &200\\ \hline
$\kappa/e$   & 3.54 $\sim$ -0.86&2.88$\sim$ -0.60&2.48 $\sim$ -0.44
&2.2 $\sim$ -0.34
&2.0$\sim$-0.28\\ \hline
$\lambda/e$ &-9.28$\sim$2.28&-7.98 $\sim$ 1.68&-7.2$\sim$1.32&-6.7
$\sim$1.08&-6.34$\sim$0.9\\ \hline
$|\tilde \kappa|/e$ &$<$1.76&$<$1.32&$<$1.06&$<$0.88&$<$0.76\\ \hline
$|\tilde \lambda|/e$ &$<$4.60&$<$3.66&$<$3.08&$<$2.68&$<$2.4\\
\end{tabular}
\label{table1}
\end{table}

\begin{table}
\caption{$Br(b\rightarrow s\gamma)$ vs. $\kappa$ for $m_t = 150 GeV$.}
\begin{tabular}{|c|c|c|c|c|c|c|c|c|}
$\kappa/e$&3.0&2.5&2.0&1.5&0.5&0.0&-0.5&-1.0\\ \hline
$Br\times 10^{4}$&9.78&5.47&2.4&0.58&7$\times 10^{-4}$& 2.57&5.72&10\\
\end{tabular}
\label{table2}
\end{table}

\begin{table}
\caption{$Br(b\rightarrow s\gamma)$ vs. $\lambda$ for $m_t=150 GeV$.}
\begin{tabular}{|c|c|c|c|c|c|c|c|}
$\lambda/e$&1.5& 0.0&-1.5&-3.0&-4.5&-6.0&-7.5\\ \hline
$Br\times10^{4}$&5.58&2.57&0.62&8$\times10^{-4}$& 0.71&2.76&6.13\\
\end{tabular}
\label{table3}
\end{table}
\end{document}